\documentclass[12pt]{article}

\usepackage{scicite}

\usepackage{times}
\usepackage{graphicx}
\usepackage{amsmath}
\usepackage{amssymb}

\newenvironment{sciabstract}{%
\begin{quote} \bf}
{\end{quote}}

\newcounter{lastnote}
\newenvironment{scilastnote}{%
\setcounter{lastnote}{\value{enumiv}}%
\addtocounter{lastnote}{+1}%
\begin{list}%
{\arabic{lastnote}.}
{\setlength{\leftmargin}{.22in}}
{\setlength{\labelsep}{.5em}}}
{\end{list}}

\title{Nonlinear $\pi$ phase shift for single fiber-guided photons interacting with a single atom}

\author{
J\"urgen Volz, Michael Scheucher, Christian Junge \& Arno Rauschenbeutel\\
\\
\normalsize{Vienna Center for Quantum Science and Technology,}\\
\normalsize{Atominstitut, Vienna University of Technology, Vienna, Austria.}
\\
\normalsize{$^\ast$To whom correspondence should be addressed; E-mail:  arno.rauschenbeutel@ati.ac.at.}
}

\date{}

\begin{document} 


\baselineskip24pt


\maketitle


\begin{sciabstract}
Realizing a strong interaction between individual optical photons is an important objective of research in quantum science and technology. Since photons do not interact directly, this goal requires, e.g., an optical medium in which the light experiences a phase shift that depends nonlinearly on the photon number. Once the additional phase shift for two photons reaches $\pi$, such an ultra-strong nonlinearity could even enable the direct implementation of high-fidelity quantum logic  operations\cite{Chuang2000}. However, the nonlinear response of standard optical media is many orders of magnitude too weak for this task\cite{Matsuda2009}. Here, we demonstrate the realization of an optical fiber-based nonlinearity that leads to an additional two-photon phase shift close to the ideal value of $\pi$. Our scheme employs a whispering-gallery-mode resonator, interfaced by an optical nanofiber, where the presence of a single rubidium atom in the resonator results in a strongly nonlinear response.\cite{Hofmann2003} We experimentally show that this results in entanglement of initially independent incident photons. The demonstration of this ultra-strong nonlinearity in a fiber-integrated system is a decisive step towards scalable quantum logics with optical photons. 
\end{sciabstract}

Optical photons are a key ingredient for investigations and applications in modern quantum information science. They are well decoupled from their environment and can, nevertheless, be conveniently manipulated with high precision. In conjunction with the ease of transmitting them over long distances using optical fibers, this makes photons prime candidates for the distribution\cite{Ritter2012} 
and processing of quantum information in scalable quantum networks\cite{Kimble2008} as well as for metrology beyond the standard quantum limit\cite{Giovannetti2004a}. 
These and many other applications require the photons to interact with each other in order to  prepare and probe entanglement or to perform quantum logic operations. However, a direct photon--photon interaction does not exist in free space. It has been shown that an effective photon--photon interaction can be implemented probabilistically using linear optical elements in combination with projective measurements.\cite{Knill2001} Yet, for now, this technique cannot considered to be scalable in a practical sense because it requires high fidelity single photon sources and detectors which still go beyond present capabilities.\cite{Kok2007}
Alternatively, a deterministic interaction can be realized by means of an optical medium that exhibits a nonlinearity down to the level of individual photons. Such strong optical nonlinearities have been demonstrated, e.g., in atomic ensembles where the nonlinearity either stems from direct interaction between the excited atoms\cite{Peyronel2012,Firstenberg2013} or from the generation of an intrinsic Kerr nonlinearity via electromagnetically induced transparency.\cite{Bajcsy2009,Lee2012} An alternative approach is based on enhancing the  nonlinearity of individual quantum emitters by coupling them to optical resonators. With these systems, nonlinear phase shifts up to a few ten degrees, single-photon controlled on- and off-switching of light, as well as photon number dependent redirection of light  have been demonstrated.\cite{Turchette1995, Fushman2008,Dayan2008,Tanji-Suzuki2011,Reinhard2012}

In this work, we demonstrate the realization of an optical fiber-based nonlinearity that leads to an unprecedented two-photon phase shift close to the ideal value of  $\pi$, following a proposal by Hofmann et al.\cite{Hofmann2003}. The nonlinearity is induced by a single $^{85}$Rb atom that is strongly coupled to a whispering-gallery-mode (WGM) resonator while the latter is efficiently interfaced by an optical nanofiber.
Our experimental setup is sketched in Fig. 1. The central element is a so-called bottle microresonator,\cite{Sumetsky2004,Lou05,Poellinger2009} a novel type of WGM microresonator which is conceptually similar to other WGM microresonators\cite{Braginsky1989,Armani2003} but has the additional advantage of being fully tunable\cite{Poellinger2009}. We interface the resonator with a tapered fiber coupler that comprises a nanofiber waist\cite{Spillane2003}. The distance between the nanofiber and the resonator is chosen such that the system operates in the overcoupled regime, i.e., that the coupling rate $\kappa_f$ between the nanofiber and the resonator is larger than the intrinsic resonator loss rate $\kappa_i$. 
In this regime, resonant, horizontally ($H$-) polarized light that is guided in the nanofiber will couple into the empty resonator and then couple back into the fiber, thereby acquiring a phase shift of $\pi$ due to the interaction with the resonator mode. The situation is different if the photon is vertically ($V$-) polarized or if an atom is strongly coupled to the resonator. In the former case, the light cannot enter due to the large birefringence of the resonator while, in the latter case, the presence of the atom changes the resonance condition. Thus, with an atom present, $H$-polarized photons that arrive one by one cannot enter the resonator mode. In both cases, the photons  therefore do no acquire a phase shift. However, due to the nonlinearity of the interaction, the system behaves differently when two $H$ photons are incident on the resonator at the same time: The presence of one photon saturates the atom--resonator system and thus modifies the resonance condition for the other photon. Since this process is coherent and the two photons are indistinguishable, this results in an additional phase shift of $\pi$ for the two-photon state compared to the case where only a single photon interacts with the atom--resonator system.
This can be measured experimentally by comparing it with the uncoupled $V$-polarized light field which acts as phase reference.
Neglecting optical losses as well as the spontaneous emission of the atom into the resonator mode (see supplementary information) and choosing an input polarization along the $+45^\circ$  (i.e., $H$+$V$) direction, the quantum state of the two photons before and after the interaction with the atom--resonator system is given by
\begin{eqnarray}
|\psi_{\rm initial}\rangle&=&
 \frac{1}{2\sqrt{2}}\left(a^\dagger_Ha^\dagger_H+2a^\dagger_Ha^\dagger_V+a^\dagger_Va^\dagger_V\right)|0\rangle=\frac{1}{\sqrt{2}}a^\dagger_Pa^\dagger_P|0\rangle \label{eqn1}
\\
|\psi_{\rm final}\rangle&=& \frac{1}{2\sqrt{2}}\left(-a^\dagger_Ha^\dagger_H+2a^\dagger_Ha^\dagger_V+a^\dagger_Va^\dagger_V\right)|0\rangle=\frac{1}{2}\left(a^\dagger_Va^\dagger_P-a^\dagger_Ha^\dagger_M\right)|0\rangle,\label{eqn2}
\end{eqnarray}
respectively. Here we introduced  the creation operators for horizontal, vertical,  plus ($P$), and minus ($M$) 45$^\circ$ polarized photons $a^\dagger_H$, $a^\dagger_V$, $a^\dagger_P=(a^\dagger_H+a^\dagger_V)/\sqrt{2}$ and $a^\dagger_M=(a^\dagger_H-a^\dagger_V)/\sqrt{2}$, respectively. Due to the nonlinear interaction, an additional phase  occurs for the $|HH\rangle$ term which, in the case of a $\pi$ phase shift, results in the minus in Eq.~(\ref{eqn2}). As a result, the final state is no longer separable.
In the experiment, optical losses, fluctuating detunings, birefringence, and spontaneous emission into the resonator mode\cite{Hofmann2003} will inevitably change the probabilities for the outcome of the individual states and lead to decoherence of their relative phase. One thus has to verify the fidelity with which the nonlinear interaction leads to the ideal output $|\psi_{\rm final}\rangle$ and if the entanglement is indeed prepared under realistic conditions.

In order to optimize our atom--resonator system, we first analyze its performance at the single photon level before investigating the nonlinear phase shift. For this purpose, we tune the bottle resonator and the incident light into resonance with the $F=3$, $m_F=3$ $\to$ $F'=4$, $m_F=4$ $D_2$-line transition of $^{85}$Rb and choose a TM-polarized resonator mode. In this case, the electric field has a dominant component perpendicular to the resonator surface and the atom cannot emit photons into the empty, counterpropagating resonator mode\cite{Junge2013}. Thus, the atom only interacts with a single resonator mode, similar to the situation in a strongly birefringent, single-sided Fabry-Perot resonator\cite{Hofmann2003}. We adjust the ratio of the amplitudes of the two polarization components, $H$ and $V$, of the incident light field such that  they have the same amplitude  after the interaction with the empty resonator. Thus, the transmitted light is $M$-polarized, thereby minimizing the power at the $P$-detector in the case when no atom is coupled. An increase in count rate on this detector then serves as a trigger that signals the arrival of an atom in the resonator mode (see supplementary material). 

First, we analyze the polarization change of the transmitted light between the empty resonator case and the case where an atom is coupled as a function of the fiber--resonator coupling strength $\kappa_f$. The latter is adjusted by changing the resonator--nanofiber distance. Without atom, the resonator imprints a phase shift close to $\pi$ on the $H$ component of the incident light and switches its polarization to $M$.  When an atom couples to the resonator, this situation changes and the phase shift disappears. From a polarization analysis, we determine the overlap of the output light field with $P$-polarization as well as the survival probability of the incident photons (see supplementary information). The results are shown in Fig. 2: One observes a monotonous increase of the photon survival probability  with the coupling rate $\kappa_f$ while the overlap with  $P$-polarization reaches a maximum of 0.82 for $\kappa_f=2\pi \times17$ MHz, in good agreement with the theoretical prediction (solid lines in Fig. 2).  In order to model this behavior, we assume a normal distribution of the atom--resonator coupling strength $g$, which originates from the motion of the atom through the resonator mode.  From the transmission properties of our atom--resonator system, we calculate the corresponding polarization change (see supplementary material). We then fit this distribution to the measured data, with the mean coupling strength, $\bar g$, and the standard deviation, $\sigma_g$, as the only free parameters. From this fit we obtain $\bar g=2\pi\times(13.5\pm1.5)$ MHz and  $\sigma_g=2\pi\times4$ MHz. 

We now choose the working point  $\kappa_f/\kappa_i=2.8$ (see dashed line in Fig. 2) at which we expect photon number-independent loss and, as a consequence, an effectively dispersive nonlinearity (see supplementary material). At this point, the atom--resonator system yields a polarization dependent transmission between 20\% for $H$- and close to 100\% for $V$-polarization. In order to experimentally determine the nonlinear phase shift, we analyze the transmitted light in three complementary polarization bases by recording coincidence counts between the different detectors (see Fig. 1). Figure 3 exemplarily shows the measured coincidences recorded for the combinations $R/L$, $R/P$, $M/M$ and $H/M$. We clearly observe photon bunching and anti-bunching for zero time delay. This indicates that two simultaneously arriving photons have a different polarization than individual photons. This could, e.g., be exploited for performing a photon number dependent routing of the fiber guided photons.\cite{Dayan2008} For comparison, we also record the correlations without an atom coupled to the resonator. There, no  significant bunching or anti-bunching is apparent. This confirms that the atom is the physical origin of the nonlinearity. From the set of independent tomographic measurements, we then reconstruct the density matrix of the two-photon state after the interaction with the cavity using a maximum likelihood estimation.\cite{James2001} Figure 4 shows the real and imaginary parts of this reconstructed density matrix $\rho_{\rm meas}$ for zero mean detection time difference and a width of 3~ns for the coincidence window. For comparison, we also show the ideal density matrix according to Eq.~(\ref{eqn2}), $\rho_{\rm final}=|\psi_{\rm final}\rangle \langle \psi_{\rm final}|$, and find good qualitative agreement. The reduction of the off-diagonal elements of $\rho_{\rm meas}$ mainly originates from the shot-to-shot fluctuations of the atomic position. This results in a fluctuating  atom--resonator coupling strength and a concomitant variation of the output state. Figure 4 also shows the overlap of the measured two-photon state with the ideal state, $\langle\psi_{\rm final}|\rho_{\rm meas}|\psi_{\rm final}\rangle$, as well as the nonlinear phase shift as a function of the mean delay between the detections of the two photons and of the width of the coincidence window. We find a maximum overlap of $0.57\pm0.01$ and a nonlinear phase shift of $(1.05\pm0.01) \pi$, close to the value of $\pi$ expected under ideal conditions.

As is apparent from Eq.~(\ref{eqn2}), the $\pi$ nonlinear phase shift ideally generates a non-separable state of light from the initially independent incident photons. In order to quantify the non-classical character of the experimentally prepared two-photon state, we calculate its concurrence\cite{Hill1997,Adamson2007}  (see supplementary material). We find a maximum concurrence of $0.27\pm0.01$ at zero mean detection time difference, thereby demonstrating that $\rho_{\rm meas}$ corresponds to a highly entangled state. The concurrence and, thus, the entanglement vanishes when the time delay between the photons increases beyond the resonator lifetime  (see Fig. 4). This is expected because, in this case, the photons interact one-by-one with the atom--resonator system.

In summary, we demonstrated an ultra-strong optical nonlinearity that leads to a nonlinear $\pi$ phase shift for coincident photons. The employed experimental platform is fully fiber-integrated and, thus, compatible with optical fiber communication networks. The scheme relies on a single atom which can, in contrast to atomic ensembles, be straightforwardly prepared a quantum mechanical superposition of two states which are coupled and not coupled to the resonator mode, respectively. This is the basis for realizing a deterministic controlled $\pi$-phase gate\cite{Duan2004} (see supplementary information), which would  enable deterministic quantum computation protocols with photons. The performance of our system can, e.g., be improved by increasing the  photon lifetime and by trapping the atom in a small volume close to the resonator, thereby avoiding the variation of the atom--resonator coupling strength. For a state-of-the-art bottle microresonator,\cite{Poellinger2009} this would result in a transmission in excess of 82\%. The system thus constitutes a powerful experimental platform for the implementation of deterministic photon--photon gates and future photonics applications like single-photon transistors\cite{Chen2013} or the deterministic generation and analysis of entangled states of light.

\bibliography{nature}

\bibliographystyle{Science}

\begin{scilastnote}
\item We gratefully acknowledge financial support by the European Science Foundation and the Volkswagen Foundation. J.V.  acknowledges support by the European Commission (Marie Curie IEF Grant 300392). C.J. acknowledges support by the German National Academic Foundation.
\end{scilastnote}

\clearpage

\includegraphics[width=0.5\textwidth]{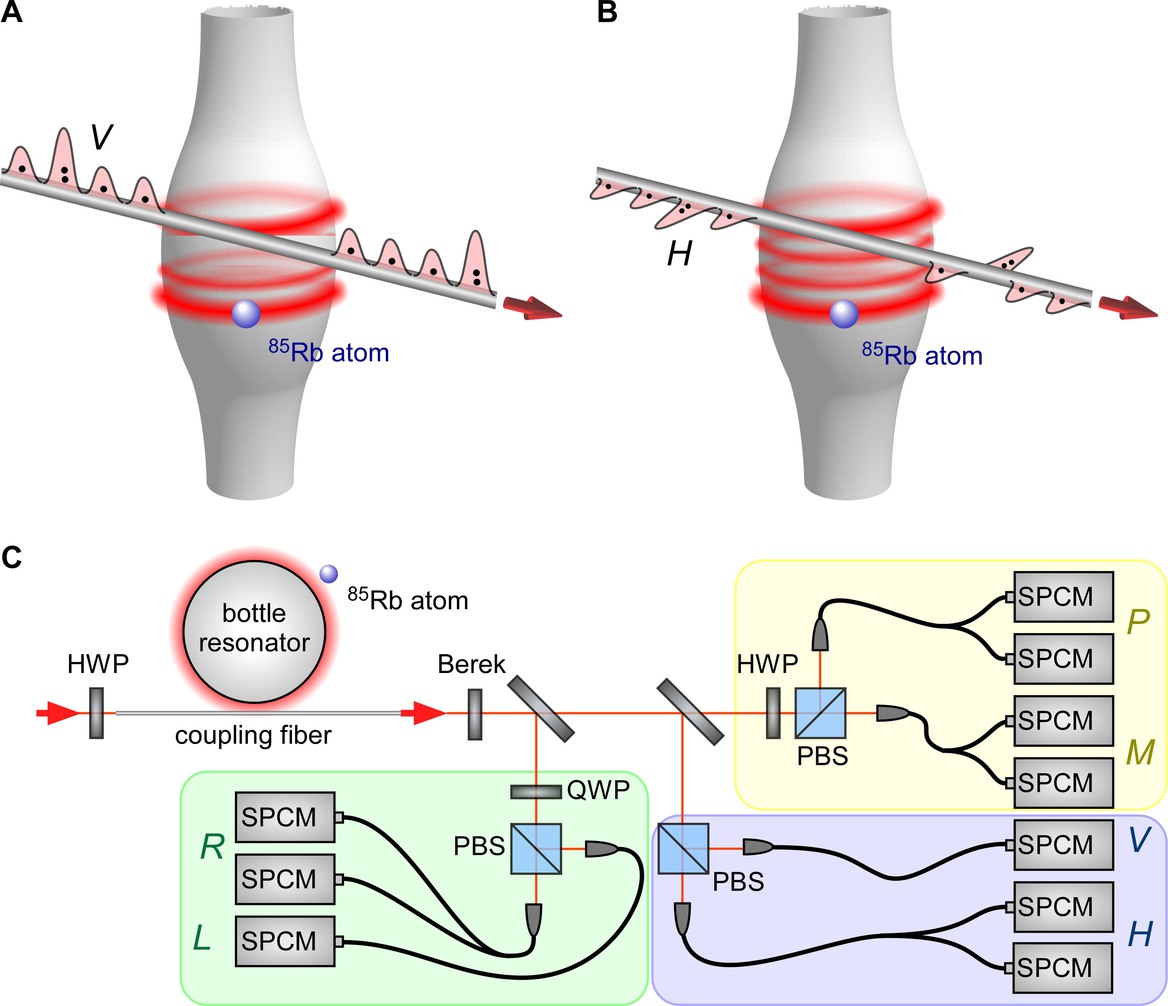}\\
\noindent {\bf Fig. 1.}
 Experimental setup showing the bottle microresonator and the coupling fiber.
\textbf{A} When the incident light is vertically ($V$-) polarized, it does not couple to the resonator mode. Thus, no phase shift occurs and the light simply passes the resonator unaffected. This polarization component serves as a phase reference. \textbf{B} When the light is horizontally ($H$-) polarized, the presence of a strongly coupled atom in the resonator prevents the light from entering and, as a consequence, no phase shift occurs. However,  in the case where two photons arrive at the same time, the response of the system changes due the nonlinearity of the atom--resonator system and an additional phase shift of $\pi$ is imprinted onto the light. \textbf{C} In order to detect the phase shift, the transmitted light is measured using a polarization analyzer, which records photon--photon correlations between single photon counting modules (SPCM) in the three complementary polarization bases $H/V$, $P/M$ and $R/L$ (right/left circularly polarized). 

\newpage
\includegraphics[width=0.5\textwidth]{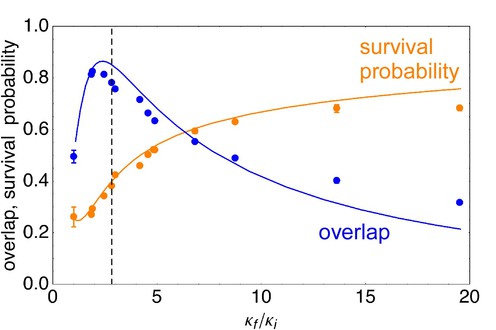}\\
\noindent {\bf Fig. 2.}
Single photon polarization change. 
Overlap of the polarization of the transmitted light with $P$-polarization and survival probability of the incident photons as a function of the resonator--fiber coupling strength $\kappa_f$ (in units of the intrinsic resonator loss rate $\kappa_i=2\pi\times8.4$ MHz). The input polarization was set such that, without an atom coupled to the resonator, the transmitted light is fully $M$-polarized. The error bars correspond to the 1$\sigma$ statistical error and the solid lines are theoretical predictions that take into account a variation of the atom--resonator coupling strength and a residual atom--resonator and atom--light detuning of $\Delta_{ar}=\Delta_{al}=2\pi\times2.2$ MHz. The maximum overlap with $P$-polarization is observed at $\kappa_f=2 \pi \times17$ MHz, in good agreement with the theoretical prediction (solid lines).
 
\newpage
\includegraphics[width=0.5\textwidth]{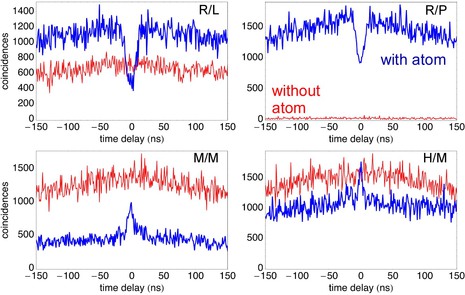}\\
\noindent {\bf Fig. 3.}
Tomographic data. Measured coincidence counts between different detectors in Fig. 1 as a function of the delay between the detection events (bin size 1 ns) for a total measurement time of 28 hours. The case where an atom couples to the resonator corresponds to the blue data. For comparison, the empty resonator case is shown in red. For clarity, we exemplarily show only the correlation functions between the detector pairs $R/L$, $R/P$, $M/M$, $H/M$ which exhibit clear photon bunching and anti-bunching. The full set of measurements including other combinations of bases is presented in the supplementary information.

\newpage
\includegraphics[width=0.8\textwidth]{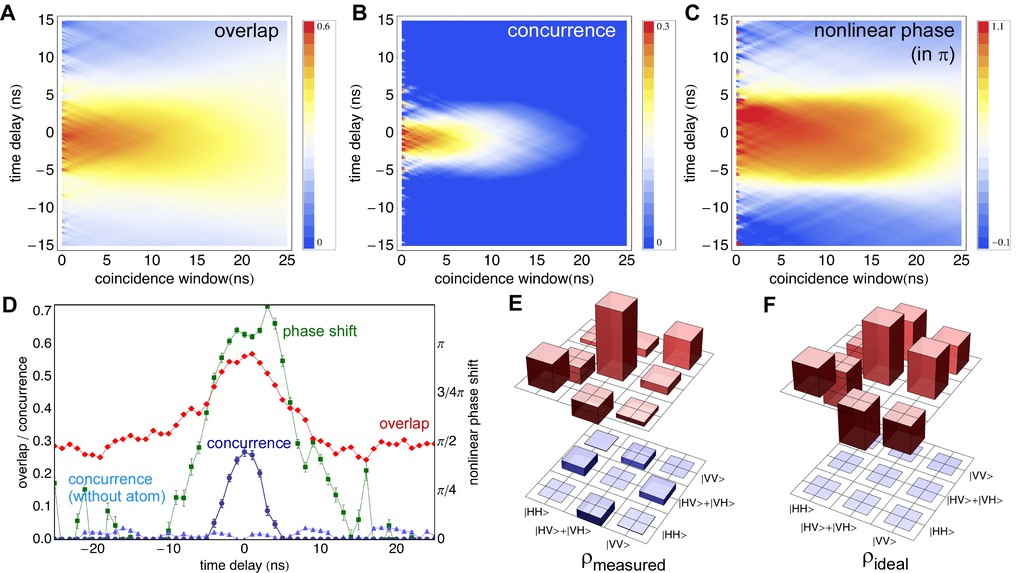}\\
\noindent {\bf Fig. 4.}
State tomography of the transmitted light. \textbf{A} Overlap of the experimentally prepared state with the ideal state $\rho_{\rm final}$ as well as \textbf{B} concurrence and \textbf{C} nonlinear phase shift of the two-photon state as a function of the coincidence window size and the mean photon--photon delay. The data is calculated from the corresponding density matrices, which are obtained from a maximum likelihood estimation. \textbf{D} Overlap, concurrence, and nonlinear phase shift for zero time delay. For reference, the concurrence in the case where no atom couples to the resonator is also shown. The error bars correspond to the 1$\sigma$ statistical error (see supplementary material). \textbf{E} Real (red) and imaginary part (blue) of  the reconstructed density matrix for zero time delay and \textbf{F} of the density matrix  $\rho_{\rm final}$ of the ideal state.

\clearpage

\section*{Supplementary Material}

\subsection*{Atom detection and experimental sequence.}
In order to couple atoms to the resonator, we use an atomic fountain that delivers a cloud of around $5\times10^7$ laser-cooled atoms to the resonator. The arrival of a single atom in the resonator mode is detected using an extension of the technique presented in Ref. [\cite{Junge2013}]. The latter allows us to detect atoms in the overcoupled regime where the fibre transmission is only marginally affected by the presence of the atom. For this purpose, we adjust the input polarisation such that the light is fully $M$-polarized after the interaction with the empty resonator. We permanently monitor the count rate of the $P$-detector and  observe an increase by two orders of magnitude when an atom enters the resonator mode. Using an field programmable gate array-based real-time detection and control system, we react to this increasing count rate within approximately 150 ns  and subsequently execute our measurement sequence during which we apply a 0.5-$\mu$s measurement light pulse. A final 1-$\mu$s probing interval ensures that the atom is still coupled to the resonator mode at the end of the measurement sequence.

\subsection*{Single photon polarisation change}
For the measurement of the polarisation change of the light that is induced by the presence of the atom, we employ the following procedure: We analyze the transmission through the coupling fibre in the $H/V$ and $P/M$ basis with and without an atom coupled to the resonator as a function of the resonator--fibre coupling strength $\kappa_f$ (see also supplementary information). In these bases, we expect the dominant effect.  $V$-polarized light does not couple to the resonator and the complex amplitude transmission through the coupling fiber is unity. In contrast, $H$-polarized light always couples to the resonator and  the transmission is given by
\begin{equation}
t_H=\frac{\kappa_L-\kappa_f+i\Delta_{rl}}{\kappa_L+\kappa_f+i\Delta_{rl}}.
\end{equation}
Here $\kappa_f$ is the fibre--resonator coupling strength, $\kappa_L$ the resonator loss rate, and $\Delta_{rl}$ the detuning between the resonator and the light field. This expression allows us to treat both cases --- the empty resonator and the coupled atom--resonator system  --- by introducing two loss rates: For the empty resonator, the loss rate is given by the intrinsic loss rate, $\kappa_L=\kappa_i=2\pi\times 8.4$ MHz, while for the coupled atom--resonator system in the single photon limit, the losses are  governed by an effective loss rate that also depends on the atom:
\begin{equation}
\kappa_L=\frac{g^2}{\gamma+i\Delta_{al}}+\kappa_i.
\end{equation}
Here, $\gamma$ is the atomic decay rate, $g$ the atom--resonator coupling strength, and $\Delta_{al}$ is the detuning between the atom and the resonator-mode. Using these expressions, we model the phase shift and transmission properties of the resonator and calculate the expected polarisation change of the transmitted light field.

\subsection*{Atom--resonator interaction beyond the single-photon limit}
In order to simulate the behavior of the atom--resonator system for the case with more than one photon  present, we perform a full numerical calculation of our system by solving the master equations of the pumped atom--resonator system using the approximation of a two-level atom interacting with a single resonator mode\cite{Junge2013}. In this case, the master equation is given by
\begin{eqnarray}
\frac{d\rho(t)}{dt}&=&-\frac{i}{\hbar}[H,\rho]+(\kappa_f+\kappa_i) (2b\rho b^\dagger-b^\dagger b\rho-\rho b^\dagger b)+\\
&&\gamma(2\sigma^-\rho\sigma^+-\sigma^+\sigma^-\rho-\rho\sigma^+\sigma^-)
\end{eqnarray}
 where 
 \begin{equation}
 H=\Delta_{rl}b^\dagger b+\Delta_{al}\sigma^{+}\sigma^{-} +g(b^\dagger\sigma^{-}+b\sigma^+)+\sqrt{2\kappa_f}\langle a_{H,\rm in} \rangle(b^\dagger+b)
 \end{equation}
 is the atom--resonator Hamiltonian in the rotating wave approximation and $b$ ($b^\dagger$) is the annihilation (creation) operator of a resonator photon, $\sigma^+$ ($\sigma^-$) is the atomic excitation (deexcitation) operator, $\rho$ is the atom--resonator density matrix and $\langle a_{H, \rm in} \rangle$ is the amplitude of the incident $H$-polarized fibre guided light. The final state of the light field after the interaction with the resonator is then given by the interference of the incident field with the field that is coupled out of the resonator:
 \begin{eqnarray}
 a_{H, \rm out}&=&a_{H, \rm in}-\sqrt{2\kappa_f}b\\
 a_{V, \rm out}&=&a_{V, \rm in}
 \end{eqnarray}
The input light field corresponds to a coherent state which allows us to replace the input operator $a_{H, \rm in}$ and $a_{V, \rm in}$ by their expectation value\cite{Gardiner1985} and to calculate the expectation values for the output fields. Using the measured distribution of coupling strengths, this allows us to predict the optical losses in the system and to estimate the optimal working point, where the two-photon transmission is independent of the photon time delay.
 
\subsection*{State reconstruction}
For the reconstruction of the density matrix of the two-photon state after the interaction with the resonator, we use the 19 non-trivial coincidence detections obtained from the polarisation setup. Note that we did not record coincidences for the settings $RR$ and $VV$ (see supplementary information). We define a coincidence window and sum up all two-photon events within this window. The incident photons originate from a coherent laser beam with a coherence time on the order of one $\mu$s and are thus, to a very good approximation, indistinguishable within the interaction time which is given by the resonator lifetime of $\kappa^{-1}\approx 7$ ns. As a consequence, the two-photon state is limited to the symmetric subspace of a two-qubit Hilbert space. The corresponding basis states are $|HH\rangle$, $|S\rangle=(|HV\rangle+|VH\rangle)/\sqrt{2}$, and $|VV\rangle$ and the density can be written as 
\begin{equation}
\rho=
\begin{pmatrix}
\rho_{HH,HH} & \rho_{HH,S} & \rho_{HH,VV} &0 \\
\rho_{S,HH} & \rho_{S,S} & \rho_{S,VV}  &0\\
\rho_{VV,HH} & \rho_{VV,S} & \rho_{VV,VV}&0  \\
0&0&0& \rho_{A,A}
\end{pmatrix} .
\end{equation}
The matrix element $\rho_{A,A}$ and all other matrix elements which contain contributions from the asymmetric basis state $|A\rangle=(|HV\rangle-|VH\rangle)/\sqrt{2}$ vanish\cite{James2001} and the $3\times3$ sub-matrix contains the full information on the experimentally prepared state. The $3\times3$ sub-matrix  is reconstructed by performing a quantum state tomography using a maximum likelihood estimation.\cite{Adamson2007}

\subsection*{Concurrence and nonlinear phase shift}
In order to quantify the degree of entanglement, we transform the full $4\times4$ density matrix into an unentangeld basis defined by the basis states $|HH\rangle$, $|HV\rangle$, $|VH\rangle$ and $|VV\rangle$ and subsequently calculate the concurrence\cite{Hill1997}. The nonlinear phase shift is determined directly from the measured $3\times3$ density matrix according to $\phi_{nl}\equiv\phi_{HH}-2\phi_{HV}=\arg(\rho_{HH,HV})-\arg(\rho_{HV,VV})$, where these two density matrix elements were chosen due to their higher signal to noise ratio.

\subsection*{Error estimation}
The statistical errors of the measured quantities were determined by adding Poissonian noise to the measured coincidences, followed by a density matrix reconstruction and the subsequent evaluation of the concurrence, nonlinear phase shift, and  overlap with $\rho_{\rm ideal}$. For each data point in Fig. 4\textbf{D} in the main text, we generate a set of 100 random density matrices and use the resulting standard deviations as our error estimate.

\subsection*{Single photon polarization change}

 \begin{figure}[ht]
 \includegraphics[width=0.9\textwidth]{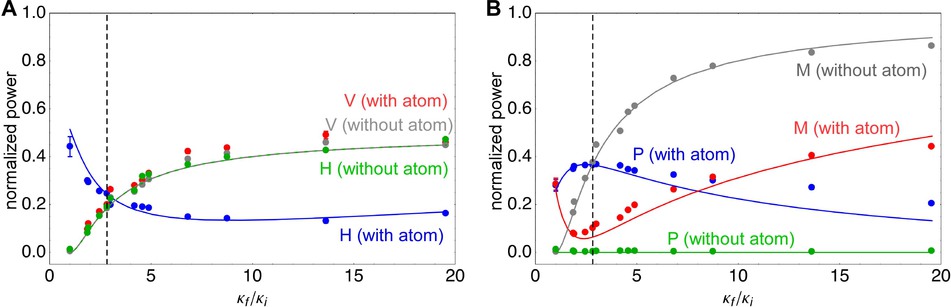}%
 \caption{\textbf{Single photon polarization change. a (b)} Transmitted power of $H$- and $V$- ($P$- and $M$-) polarized light, normalized to the total incident power (see text) as  a function of the resonator--fiber coupling strength $\kappa_f$ in units of $\kappa_i=2\pi\times8.4$ MHz. \label{Fig:S1}}
 \end{figure}

 In order to measure the effect of the resonator on the polarization of the light transmitted through the fiber, we adjust the amplitude $|\alpha_H|$ and $|\alpha_V|$ of the $H$- and $V$-polarization component such that after the interaction with the empty resonator $|t_{H,0}\cdot\alpha_H|=|\alpha_V|$ where $t_{H,0}$ is the amplitude transmission coefficient for $H$-polarized light.  $V$-polarized light  does not couple to the resonator and, thus, its transmission coefficient is unity.  The total incident power is $I_0=|\alpha_H|^2+|\alpha_V|^2$ and the relative phase between $H$ and $V$ is chosen such that after the $\pi$ phase shift of the empty resonator, the transmitted light is $M$-polarized. As a consequence, the transmission to the $P$ detector is zero (see Fig. \ref{Fig:S1}\textbf{B}) and we observe the same transmitted power $|t_{H,0}\cdot\alpha_H|^2/I_0$ and $|\alpha_V|^2/I_0$ for the $H$ and $V$ component (see Fig.~\ref{Fig:S1}\textbf{A}) which approaches $0.5I_0$ for large $\kappa_f$ due to the vanishing optical losses, $|t_{H,0}|\to1$.

When an atom couples to the resonator, the absolute value and the phase of the amplitude transmission of the $H$-polarized light changes to $t_{H,A}$ while it stays unity for $V$-polarization. Figure~\ref{Fig:S1}\textbf{A} shows that the transmitted power $|t_{H,A}\cdot\alpha_H|^2/I_0$ decreases with increasing nanofiber--resonator coupling strength $\kappa_f$. In the $P/M$ basis (see Fig.~\ref{Fig:S1}\textbf{B}) one observes an increase in the transmitted power $|t_{P,A}\cdot\alpha_P|^2/I_0$  to the $P$ detector which reaches maximum at the point where $|t_{H,A}|=|t_{H,0}|$, and then drops for larger $\kappa_f$. The transmitted power  $|t_{M,A}\cdot\alpha_M|^2/I_0$ to the $M$-detector shows the opposite behavior. 
 The solid lines are fits as described in the main text. From this measurement, we obtain the data shown in Fig. 2 of the main text.\\



\subsection*{Set of measured coincidences} 
\begin{figure}[ht]
 \includegraphics[width=0.62\textwidth]{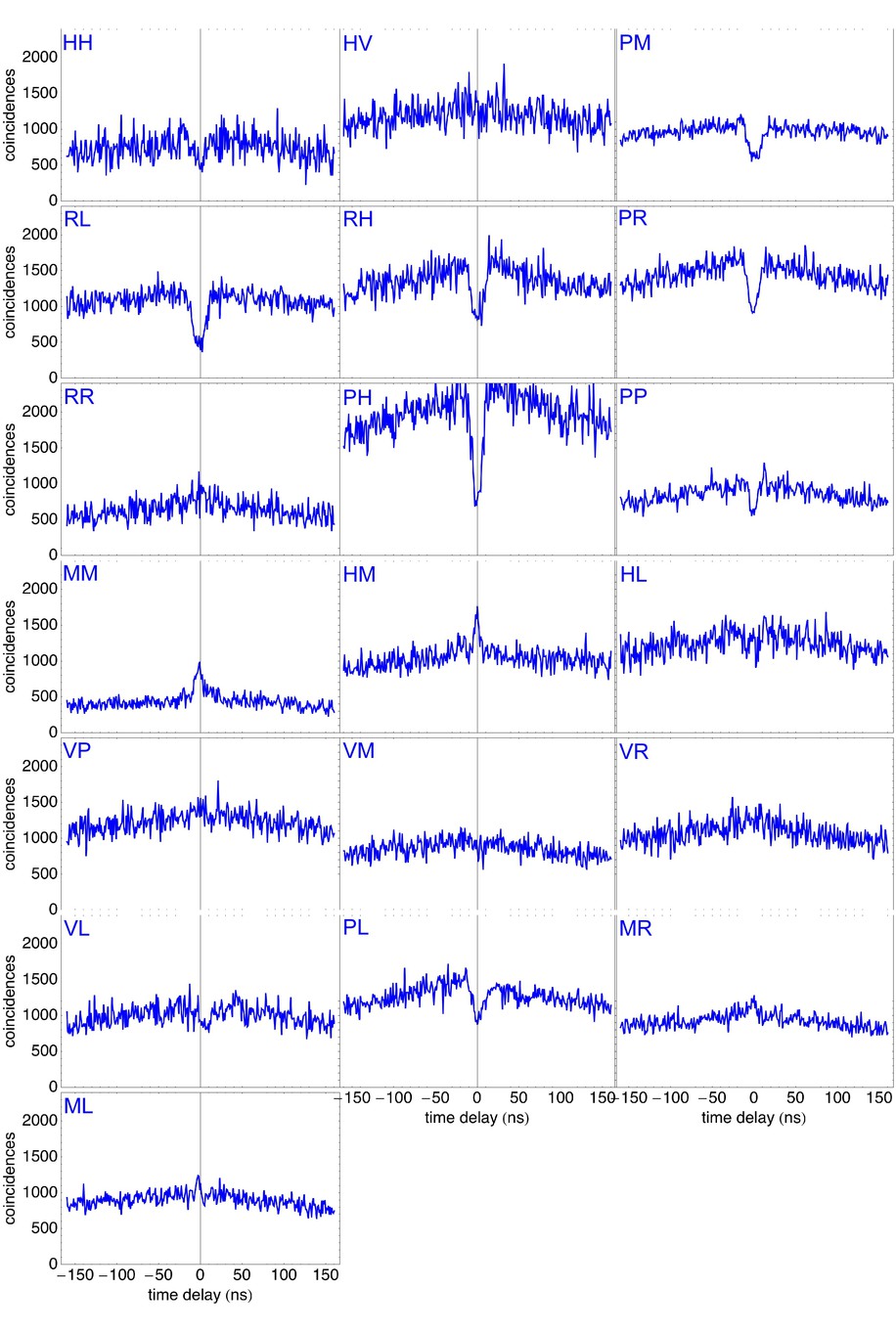}%
 \caption{\textbf{Measured coincidences}. The respective combination of bases is indicated in the upper left corner of each plot. \label{Fig:S2}}
 \end{figure}

Using the setup shown in Fig. 1 of the main manuscript, we measure 19 non-trivial coincidences. The coincidence settings $RR$ and $VV$  are not accessible due to the limited number of single photon counting modules. The full set of measured coincidences is shown in Fig. \ref{Fig:S2}.
\clearpage

\subsection*{Influence of spontaneous emission}
For clarity, we used a strongly simplified description of the nonlinear interaction of the photons with the atom--resonator system in the main text. In particular, we neglected the spontaneous emission of the atom into the resonator mode. However, as described in a semiclassical model by Hofmann et al.\cite{Hofmann2003_s}, this spontaneous emission has important physical implications for the implementation of the optical nonlinearity. For the nonlinear interaction demonstrated in our manuscript, there exist two regimes, where either a single or two photons interact with the resonator at the same time. As shown in,\cite{Hofmann2003_s} the contribution of the spontaneous emission cannot be made negligible for both regimes simultaneously. Since photon which are spontaneously emitted into the resonator mode have a random phase, this leads to a spread of the phase distribution and thus to decoherence of the final state of the light in Eq.~(2) in the main text. 
Depending on the application, this spread of the phase distribution can have important consequences. In particular, it prevents the direct use of the demonstrated Kerr nonlinearity for the realization of a high-fidelity photon--photon quantum gate. This issue can, however, be circumvented by using a two-step interaction process (see following section).
We note that a quantitative analysis of the influence of the quantum noise due to spontaneous emission on the performance of the demonstrated optical nonlinearity requires a fully quantum mechanical treatment of the spatiotemporal one- and two-photon wavefunctions.\cite{Kojima2003_s}

\subsection*{Implementation of a photon-photon sign-flip gate}
While the nonlinearity demonstrated in our work cannot directly be exploited for the realization of a high-fidelity photon--photon gate,\cite{Hofmann2003_s,Shapiro2006_s} a deterministic controlled sign-flip gate for photons can be realized by storing the first photon in the atom--resonator system prior to the interaction with the second photon: We assume that the mode of incident photons is defined by their temporal degree of freedom, i.e., the single photon pulses have a relative delay which is much larger than their coherence time. The atom is initially prepared in a state $|g_{\rm coupl}\rangle$ that couples to the resonator (e.g., the $F=3$ hyperfine ground state of $^{85}$Rb). We consider the case where the two incident photons are $P$-polarized. Thus, the total state before the interaction with the resonator is given by
\begin{equation}
|\Psi\rangle=\frac{1}{2}(|H_2\rangle+|V_2\rangle)(|H_1\rangle+|V_1\rangle)|g_{\rm coupl}\rangle
\end{equation}
When the first photon arrives at the resonator we apply an additional control laser which performs a Raman transition of the atom from $|g_{\rm coupl}\rangle$ to $|g_{\rm uncoupl}\rangle$ if and only if the incident photon $1$ is $H$-polarized and enters the resonator. Here, $|g_{\rm uncoupl}|\rangle$ is a second ground state of the atom that does not couple to the resonator (e.g., the $F=2$ hyperfine ground state of $^{85}$Rb). As a consequence, the $H$-photon is stored in the atom and the state is now given by
\begin{equation}
|\Psi\rangle=\frac{1}{2}(|H_2\rangle+|V_2\rangle)(|0\rangle|g_{\rm uncoupl}\rangle+|V_1\rangle|g_{\rm coupl}\rangle)
\end{equation}
Thus, if the first photon is $H$ polarized, the atom is switched to a state that is off-resonant with respect to the resonator mode and the $H$ component of the second photon will acquire a phase shift of $\pi$. Otherwise, the atom is resonant with the resonator mode and no phase shift occurs. The total state after the interaction of the second photon with the resonator thus reads
\begin{equation}
|\Psi\rangle=\frac{1}{2}(-|H_2\rangle|0\rangle|g_{\rm uncoupl}\rangle+|V_2\rangle|0\rangle|g_{\rm uncoupl}\rangle+|H_2\rangle|V_1\rangle|g_{\rm coupl}\rangle+|V_2\rangle|V_1\rangle|g_{\rm coupl}\rangle).
\end{equation}
In the next step, the stored photon can be released using a second Raman process and ones end up in the final state
\begin{equation}
|\Psi\rangle=\frac{1}{2}\left(-|H_2\rangle|H_1\rangle+|V_2\rangle|H_1\rangle+|H_2\rangle|V_1\rangle+|V_2\rangle|V_1\rangle\right)|g_{\rm coupl}\rangle,
\end{equation}
which is the desired state. We note that other implementations of a controlled phase-flip gate which do not require the implementation of Raman transitions have been proposed.\cite{Duan2004_s}

\end{document}